\documentclass[twocolumn,showpacs,superscriptaddress,10pt]{revtex4-1}

\usepackage[dvipdfmx,dvipdfmx]{graphicx}
\usepackage{graphicx}
\usepackage{color}
\usepackage{siunitx}
\usepackage{amssymb}

\usepackage{subfigure}

\usepackage{epsfig}
\usepackage{float}
\usepackage{epstopdf}

\begin{document}

\title{Nitrogen-vacancy defect emission spectra in the vicinity of an adjustable silver mirror}


\author{Niels M. Israelsen}
\email{nikr@fotonik.dtu.dk}
\affiliation{Center for Macroscopic Quantum States (bigQ), Department of Physics, Technical University of Denmark, Building 307, Fysikvej, 2800 Kgs. Lyngby, Denmark}
\affiliation{DTU Fotonik, Department of Photonics Engineering, Technical University of Denmark, {\O}rsteds Plads Building 343, 2800 Kongens Lyngby, Denmark}

\author{Ilya P. Radko}
\affiliation{Center for Macroscopic Quantum States (bigQ), Department of Physics, Technical University of Denmark, Building 307, Fysikvej, 2800 Kgs. Lyngby, Denmark}

\author{Alexander Huck}
\email{alexander.huck@fysik.dtu.dk}
\affiliation{Center for Macroscopic Quantum States (bigQ), Department of Physics, Technical University of Denmark, Building 307, Fysikvej, 2800 Kgs. Lyngby, Denmark}

\author{Ulrik L. Andersen}
\affiliation{Center for Macroscopic Quantum States (bigQ), Department of Physics, Technical University of Denmark, Building 307, Fysikvej, 2800 Kgs. Lyngby, Denmark}

\begin{abstract}
Optical emitters of quantum radiation in the solid state are important building blocks for emerging technologies making use of the laws of quantum mechanics. The efficiency of photon extraction from the host material is low for many solid-state systems due to their relatively high index of refraction. In this article we experimentally study the emission spectrum of an ensemble of nitrogen-vacancy defects implanted around 8~nm below the planar diamond surface and in the vicinity of a planar silver mirror. Scanning the distance between diamond and the mirror, we observe an enhancement of the spectral emission power by up to a factor of 3. We construct a model based on classical dipoles and elucidate the observations as being caused by interference in the far field of the emitters.
\end{abstract}
\maketitle

The nitrogen-vacancy (NV) defect in diamond has received significant attention in the past two decades, mainly due to the excellent coherence properties of the associated electron spin~\cite{Balasubramanian2009}, the optical spin polarization and readout techniques~\cite{Jelezko2004}, and the possibility of emitting single photons at room temperature~\cite{Kurtsiefer2000}. With these properties, the NV defect has emerged as a prime candidate for novel quantum technologies with a focus on applications in quantum information processing~\cite{Dutt2007} and sensing of magnetic fields~\cite{Taylor2008,Maze2008,Balasubramanian2008}, electric fields~\cite{Dolde2011} and temperature~\cite{Acosta2010}. Moreover, by mapping the polarization of an electron spin to nearby nuclear spins may eventually facilitate the construction of a quantum simulator~\cite{Cai2013}.

Spin-photon entanglement~\cite{Togan2010} and quantum interference of photons from two independent NV defects~\cite{Bernien2012, Sipahigil2012} has been observed. Such features are essential for the demonstrations of quantum teleportation~\cite{Bernien2013} and entanglement of spins separated by a large distance~\cite{Hensen2015}. The success rate of these protocols among many other quantities is strictly limited by the collection efficiency of indistinguishable photons within the zero-phonon line (ZPL) of the NV center. At cryogenic temperatures, only about 3\% of photons are emitted into the ZPL and typically $\ll1\%$ of photons are detected with the collection optics of the setup. Enhancing the emission rate or increasing the photon collection efficiency from the far field have been suggested to increase the attainable photon counts on the detector, and demonstrations include broadband plasmonic enhancement~\cite{Huck2011}, plasmon-based cavities~\cite{Leon2012}, an open micro-cavity system~\cite{Riedel2017}, or solid immersion lenses milled into the host material~\cite{Hadden2010}. However, these approaches are technically challenging in terms of device processing, nano-fabrication, and control.  



In this Letter, we investigate the spectral emission properties of NV defects in a bulk diamond sample and in proximity to a planar reflecting interface with adjustable distance. Our experiment shows that the collection efficiency in an adjustable spectral range from NV defects implanted approximately 8~nm below the surface can be increased by up to a factor of 3. Our experimental findings are explained by a modification of the NV center far-field emission pattern and collection efficiency into the numerical aperture (NA) of our collection optics.

\begin{figure}[tbh]
\centering{\includegraphics[width=1\linewidth]{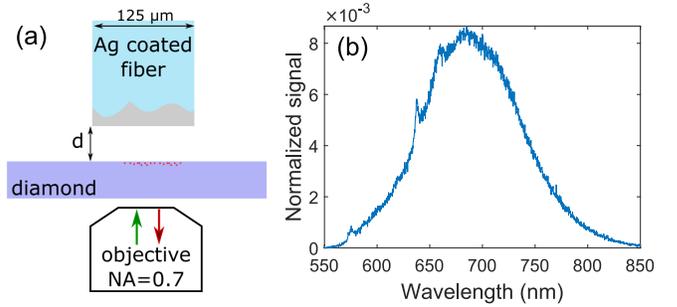} }
\caption{(a) Schematic illustration of the experimental setup: the diamond sample with a thickness of $300\,\mu$m and NV centers implanted 8~nm below the surface is mounted in a standard confocal setup equipped with a NA=0.7 microscope objective. A cleaved optical fiber coated with silver is mounted on top of the diamond with a piezo actuator used to adjust the distance $d$ to the diamond sample. (b) Reference NV emission spectrum $S_{ref,0}(\lambda)$ normalized to unit emission power and representative for the diamond sample. \label{Fig1}}
\end{figure}

Our experimental setup is illustrated in Fig.~\ref{Fig1}(a). We use an electronic-grade bulk diamond sample from Element Six with a thickness of 300~$\mu$m. Nitrogen atoms are implanted with an energy of 5~keV and a dose of $10^{13}~\mathrm{cm}^{-2}$ approximately 8~nm below the diamond surface. The sample is subsequently annealed at 800$^{\circ}$C in high vacuum yielding a uniform distribution of NV defects. The optical investigations are carried out in a standard confocal microscope, equipped with a 532-nm continuous-wave laser for the optical excitation of NV defects, an avalanche photo diode for fluorescence detection, and a spectrometer with 500~mm focal length and a grating with 150~lines/mm for spectral analysis. We use an objective lens with a NA of 0.7 and an adjustable collar (Olympus LUCPLFLN 60x) to partially correct for aberrations in the diamond sample. The reflecting interface atop the diamond sample is made of a cleaved optical fiber with a diameter of 125~$\mu$m and coated with a more than 300-nm-thick layer of silver deposited using thermal evaporation. The fiber is mounted on a piezo-actuator stage with a fine-tuning range of 20~$\mu$m to adjust the distance $d$ between the diamond surface and the fiber mirror with nm precision.

We begin the optical investigations by recording a reference fluorescence spectrum $S_{ref}$ from the NV layer without the fiber mirror. The result is shown in Fig.~\ref{Fig1}(b). The spectrometer covers the spectral range 540--900~nm, which includes the ZPL of the neutral (NV$^0$) and the negative (NV$^-$) charge state of the NV defect and their phonon side-bands, respectively. The optical excitation power was chosen to be about half the saturation pump power of single NV$^-$ defects in our setup in similar diamond with lower NV concentration, and held constant for all investigations reported in this article. We also recorded spectra from 10 different locations away from the reference point and along a line in steps of 1~$\mu$m. These spectra are nearly identical to each other, which confirms the uniformity of NV defects in the implantation region. They all show a noticeable ZPL of NV$^-$ at around 637~nm and a broad contribution due to scattering on lattice phonons in the range 640--800~nm. The tiny peak at 575~nm indicates a contribution of NV$^0$ and hence small fluctuations between the two possible charge states. 

\begin{figure}[tbh]
\centering{\includegraphics[width=1.0\linewidth]{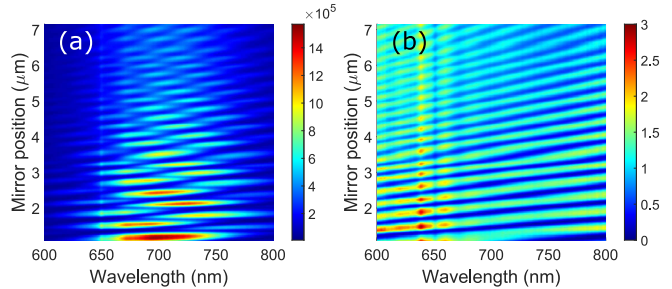}}
\caption{(a) Absolute NV defect emission spectra and (b) enhancement spectra recorded as a function of mirror position $d$ relative to the diamond sample surface. See the main text for details. \label{Fig2}}
\end{figure}

Next, we mounted the silver-coated fiber in the confocal setup, centered it on top of the optical excitation spot and minimized the distance to the diamond sample. Since the fiber cleaving process did not yield an absolutely flat surface, irregularities of the fiber surface were seen when touching the diamond at different points and thus the smallest possible distance $d_0$ we could achieve was in the range from 0.5~$\mu$m to 1~$\mu$m. Keeping the excitation at a fixed reference point on the diamond and increasing $d$ in steps of 10~nm, we recorded NV center emission spectra $S(d,\lambda)$ for each position of the fiber (here $\lambda$ is the emission wavelength). The obtained spectra are plotted as a heat map in Fig.~\ref{Fig2}(a); they produce a complex pattern which is caused by the interference of both the pump laser and NV center fluorescence. Interference of the pump laser solely modulates the pump power at the location of the NV centers, but does not impact the NV emission spectrum $S(d,\lambda)$ for a given distance~$d$. In order to remove the effect caused by laser interference, we normalized the spectra for each distance $d$ to unit spectral counts, $\int S_0(d,\lambda) d\lambda = 1$. Finally, the spectral enhancement $E$ is obtained by comparing the normalized emission spectra to the normalized reference spectrum in which the mirror is removed from the setup, $E=S_0(d,\lambda)/S_{ref,0}(\lambda).$ The distance-resolved enhancement spectrum $E$ is plotted in Fig.~\ref{Fig2}(b) and the overall structure resembles the interference pattern of an optical resonator. It is clear that when increasing the distance $d$, the number of eigenmodes in wavelength space increases. An irregularity in the pattern observed for a mirror position $d\approx 2~\mu$m we attribute to a slow thermal drift of the fiber relative to the diamond and it occurs due to the long total measurement time of around 8~h.   

In our model we consider the ensemble of NV defects as non-interacting and classical electric point dipoles with angular frequency $\omega = 2\pi c/\lambda$ located close to an interface with planar stratified media, where $c$ is the speed of light in vacuum. The radiation dynamics~\cite{Lukosz1977} and the radiation pattern~\cite{Lukosz1979} of electric dipoles in such a configuration have both been described in detail by Lukosz and Kunz. Here we apply their model to calculate the power radiated by an ensemble of electric dipoles close to stratified media into the NA of the microscope objective. The stratified media in our case are formed by the diamond with planar surfaces, an air gap of adjustable size $d$, and the fiber end-facet coated with silver. For a set of $d$ and $\lambda$, and separately for dipoles parallel ($||$) and perpendicular ($\perp$) to the diamond-air interface, we first calculate the total radiation strength $\Gamma(d,\lambda)$ relative to the total radiation strength of a similar dipole in bulk diamond $\Gamma_0$. The power detected $P(d,\lambda)$ is obtained by integrating the angular power density over the radiation modes within the NA of the objective. The theoretical enhancement spectrum is finally obtained by normalizing $P(d,\lambda)$ to the collected power when the mirror is removed from the setup.

In the calculation, we considered two classical dipoles for the emission of a NV defect with the dipole being oriented in a plane perpendicular to the NV symmetry axis~\cite{Kaiser2009}. Here, we assume that the orientation of the dipoles is homogeneously distributed in the plane. Averaging over the four possible orientations of NV defects in the diamond crystal lattice results in the relative dipole strength parallel $a_{||} \approx 0.659$ and perpendicular $a_{\perp} \approx 0.341$ to the planar diamond-air interface, which nearly is a homogeneous distribution of dipoles. The refractive index of silver over the entire spectral range $n_{Ag} (\lambda)$ is taken from Johnson and Christy~\cite{Johnson1974}, while the refractive index of diamond is assumed to be constant with $n_{d} = 2.41$.

\begin{figure}[tbh]
\centering{\includegraphics[width=1\linewidth]{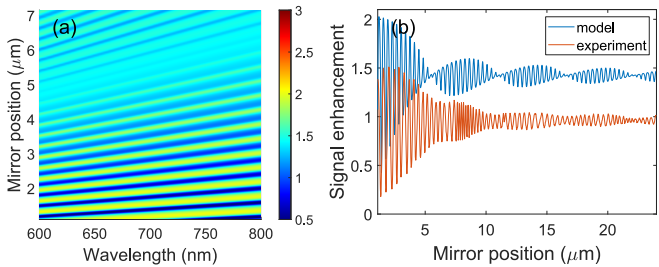}}
\caption{(a) Calculated enhancement spectrum and (b) line cuts comparing the calculated and measured enhancement spectrum at $\lambda=700$~nm. \label{Fig3}}
\end{figure}

The calculated enhancement spectrum for the same range of mirror position and $\lambda$ as in the experiment is presented in Fig.~\ref{Fig3}(a). We obtained qualitatively the best match with the experiment when limiting the NA to 0.35 in the calculation. This observation we attribute to aberrations due to the high index of refraction and dispersion of diamond, which was not accounted for by the objective. Comparing the number of oscillations within the mirror scanning range (for a fixed $\lambda$) between experiment and simulation as well as the slope of the first mode when the mirror is closest to the diamond, we can estimate the smallest mirror distance $d_0$ in the experiment to be around 0.5~$\mu$m.

Line cuts of the enhancement spectrum for $\lambda=700$~nm taken from the experiment and simulation are presented in Fig.~\ref{Fig3}(b). While the model suggests an enhancement by around 2, a factor of up to around 1.5 is observed for this wavelength in the experiment for mirror position $d<5~\mu$m. Furthermore, the feature occurring as a low-frequency beating (with nodes at $d\approx5.5$, 11, 16.5, and $22~\mu$m) in the theoretical simulation is less pronounced experimentally. This subtle difference we attribute to the potential misalignment of the silver mirror with respect to the diamond surface; the mirror was likely not perfectly parallel to the diamond surface which was not accounted for in our model. We would like to note that we use a planar geometry in modelling, i.e., the mirror has an infinite size and hence maintains its effect even at infinite distance. In the experiment and due to the limited lateral mirror size, final tilt, and surface roughness, retracting the mirror to infinity effectively excludes it from the experiment. This explains the different limits for the signal enhancement at infinite distance in the experiment and in our model. Furthermore, we would like to emphasize that our observations are not due to a Purcell enhancement. In fact, for our configuration we calculate a moderate Purcell factor in the range of  0.68--0.75 when the mirror is more than 100~nm away from the diamond. At shorter distances, there is larger Purcell enhancement, but there is no spectral enhancement due to quenching of radiation near metallic surface of the mirror.

\begin{figure}[tbh]
\centering{\includegraphics[width=1\linewidth]{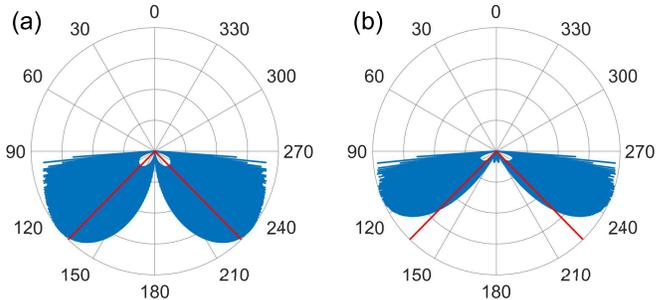}}
\caption{Calculated angular power density for (a) $d=130$~nm and (b) $d=350$~nm, while $\lambda=700$~nm. \label{Fig4}}
\end{figure}

Instead, our experimental observations are well explained by interference and a modification of the emitter far-field emission pattern depending on the mirror position $d$. In Fig.~\ref{Fig4}, we compare the calculated angular power density (at $\lambda=700$~nm) for $d=130$~nm and $d=350$~nm, which represents cases of $E>1$ and $E<1$, respectively. Both the displayed radiation pattern as well as the power radiated within the NA of our setup indicated by solid lines differ significantly. It is interference in the far field of the classical dipoles which causes the change and therefore explains the observation.

Finally, we note that in the experiment and for $d\lessapprox5~\mu$m, the enhancement at $\lambda=637$~nm and $\lambda=659$~nm is significantly more pronounced than for other wavelengths -- an enhancement factor of up to 3 is observed [cf. Fig.~\ref{Fig2}(b)]. Our model based on classical dipoles with an emission $\lambda$ clearly does not explain such features in the emission spectrum. These features are observed at the NV$^-$ ZPL and the first-order phonon scattering peak red shifted by $\approx65$~meV, which might indicate complex emitter dynamics requiring a more sophisticated model including quantized energy levels and scattering on lattice phonons. 

In conclusion, we experimentally studied the emission spectrum of NV defects implanted in a bulk diamond sample near its surface and in the vicinity of a planar silver-coated mirror. We experimentally observe a spectral enhancement by up to a factor of 3, which can be tuned in spectrum by adjusting the distance between the diamond and the mirror. We constructed a simple model with classical dipoles and attribute our observations to interference in the far field of the emitter radiation. Enhancement features at the NV$^-$ ZPL and the first-order phonon scattering peak are not explained by this model and require further investigations. Our experimental approach is easy to implement and may be applied to other solid-state systems such as emitters in GaAs.
\begin{acknowledgments}
We greatly acknowledge funding from the Danish Research Council through a Sapere Aude grant
(DIMS, Grant No. 4181-00505B). We thank Nicole Raatz in the group
of Jan Meijer at the University of Leipzig for controllably generating NV defects in our diamond samples via ion bombardment.
\end{acknowledgments}

\bibliography{References}

\begin{thebibliography}{23}
\expandafter\ifx\csname natexlab\endcsname\relax\def\natexlab#1{#1}\fi
\expandafter\ifx\csname bibnamefont\endcsname\relax
  \def\bibnamefont#1{#1}\fi
\expandafter\ifx\csname bibfnamefont\endcsname\relax
  \def\bibfnamefont#1{#1}\fi
\expandafter\ifx\csname citenamefont\endcsname\relax
  \def\citenamefont#1{#1}\fi
\expandafter\ifx\csname url\endcsname\relax
  \def\url#1{\texttt{#1}}\fi
\expandafter\ifx\csname urlprefix\endcsname\relax\def\urlprefix{URL }\fi
\providecommand{\bibinfo}[2]{#2}
\providecommand{\eprint}[2][]{\url{#2}}

\bibitem[{\citenamefont{Balasubramanian
  et~al.}(2009)\citenamefont{Balasubramanian, Neumann, Twitchen, Markham,
  Kolesov, Mizuochi, Isoya, Achard, Beck, Tissler
  et~al.}}]{Balasubramanian2009}
\bibinfo{author}{\bibfnamefont{G.}~\bibnamefont{Balasubramanian}},
  \bibinfo{author}{\bibfnamefont{P.}~\bibnamefont{Neumann}},
  \bibinfo{author}{\bibfnamefont{D.}~\bibnamefont{Twitchen}},
  \bibinfo{author}{\bibfnamefont{M.}~\bibnamefont{Markham}},
  \bibinfo{author}{\bibfnamefont{R.}~\bibnamefont{Kolesov}},
  \bibinfo{author}{\bibfnamefont{N.}~\bibnamefont{Mizuochi}},
  \bibinfo{author}{\bibfnamefont{J.}~\bibnamefont{Isoya}},
  \bibinfo{author}{\bibfnamefont{J.}~\bibnamefont{Achard}},
  \bibinfo{author}{\bibfnamefont{J.}~\bibnamefont{Beck}},
  \bibinfo{author}{\bibfnamefont{J.}~\bibnamefont{Tissler}},
  \bibnamefont{et~al.}, \bibinfo{journal}{Nature Materials}
  \textbf{\bibinfo{volume}{8}}, \bibinfo{pages}{383} (\bibinfo{year}{2009}),
  ISSN \bibinfo{issn}{1476-1122}.

\bibitem[{\citenamefont{Jelezko et~al.}(2004)\citenamefont{Jelezko, Gaebel,
  Popa, Gruber, and Wrachtrup}}]{Jelezko2004}
\bibinfo{author}{\bibfnamefont{F.}~\bibnamefont{Jelezko}},
  \bibinfo{author}{\bibfnamefont{T.}~\bibnamefont{Gaebel}},
  \bibinfo{author}{\bibfnamefont{I.}~\bibnamefont{Popa}},
  \bibinfo{author}{\bibfnamefont{a.}~\bibnamefont{Gruber}}, \bibnamefont{and}
  \bibinfo{author}{\bibfnamefont{J.}~\bibnamefont{Wrachtrup}},
  \bibinfo{journal}{Physical Review Letters} \textbf{\bibinfo{volume}{92}},
  \bibinfo{pages}{1} (\bibinfo{year}{2004}), ISSN \bibinfo{issn}{0031-9007}.

\bibitem[{\citenamefont{Kurtsiefer et~al.}(2000)\citenamefont{Kurtsiefer,
  Mayer, Zarda, and Weinfurter}}]{Kurtsiefer2000}
\bibinfo{author}{\bibfnamefont{C.}~\bibnamefont{Kurtsiefer}},
  \bibinfo{author}{\bibfnamefont{S.}~\bibnamefont{Mayer}},
  \bibinfo{author}{\bibfnamefont{P.}~\bibnamefont{Zarda}}, \bibnamefont{and}
  \bibinfo{author}{\bibfnamefont{H.}~\bibnamefont{Weinfurter}},
  \bibinfo{journal}{Phys. Rev. Lett.} \textbf{\bibinfo{volume}{85}},
  \bibinfo{pages}{290} (\bibinfo{year}{2000}).

\bibitem[{\citenamefont{Dutt et~al.}(2007)\citenamefont{Dutt, Childress, Jiang,
  Togan, Maze, Jelezko, Zibrov, Hemmer, and Lukin}}]{Dutt2007}
\bibinfo{author}{\bibfnamefont{M.~V.~G.} \bibnamefont{Dutt}},
  \bibinfo{author}{\bibfnamefont{L.}~\bibnamefont{Childress}},
  \bibinfo{author}{\bibfnamefont{L.}~\bibnamefont{Jiang}},
  \bibinfo{author}{\bibfnamefont{E.}~\bibnamefont{Togan}},
  \bibinfo{author}{\bibfnamefont{J.}~\bibnamefont{Maze}},
  \bibinfo{author}{\bibfnamefont{F.}~\bibnamefont{Jelezko}},
  \bibinfo{author}{\bibfnamefont{A.~S.} \bibnamefont{Zibrov}},
  \bibinfo{author}{\bibfnamefont{P.~R.} \bibnamefont{Hemmer}},
  \bibnamefont{and} \bibinfo{author}{\bibfnamefont{M.~D.} \bibnamefont{Lukin}},
  \bibinfo{journal}{Science} \textbf{\bibinfo{volume}{316}},
  \bibinfo{pages}{1312} (\bibinfo{year}{2007}).

\bibitem[{\citenamefont{Taylor et~al.}(2008)\citenamefont{Taylor, Cappellaro,
  Childress, Jiang, Budker, Hemmer, Yacoby, Walsworth, and Lukin}}]{Taylor2008}
\bibinfo{author}{\bibfnamefont{J.~M.} \bibnamefont{Taylor}},
  \bibinfo{author}{\bibfnamefont{P.}~\bibnamefont{Cappellaro}},
  \bibinfo{author}{\bibfnamefont{L.}~\bibnamefont{Childress}},
  \bibinfo{author}{\bibfnamefont{L.}~\bibnamefont{Jiang}},
  \bibinfo{author}{\bibfnamefont{D.}~\bibnamefont{Budker}},
  \bibinfo{author}{\bibfnamefont{P.~R.} \bibnamefont{Hemmer}},
  \bibinfo{author}{\bibfnamefont{a.}~\bibnamefont{Yacoby}},
  \bibinfo{author}{\bibfnamefont{R.}~\bibnamefont{Walsworth}},
  \bibnamefont{and} \bibinfo{author}{\bibfnamefont{M.~D.} \bibnamefont{Lukin}},
  \bibinfo{journal}{Nature Physics} \textbf{\bibinfo{volume}{4}},
  \bibinfo{pages}{810} (\bibinfo{year}{2008}), ISSN \bibinfo{issn}{1745-2473}.

\bibitem[{\citenamefont{Maze et~al.}(2008)\citenamefont{Maze, Stanwix, Hodges,
  Hong, Taylor, Cappellaro, Jiang, Dutt, Togan, Zibrov et~al.}}]{Maze2008}
\bibinfo{author}{\bibfnamefont{J.~R.} \bibnamefont{Maze}},
  \bibinfo{author}{\bibfnamefont{P.~L.} \bibnamefont{Stanwix}},
  \bibinfo{author}{\bibfnamefont{J.~S.} \bibnamefont{Hodges}},
  \bibinfo{author}{\bibfnamefont{S.}~\bibnamefont{Hong}},
  \bibinfo{author}{\bibfnamefont{J.~M.} \bibnamefont{Taylor}},
  \bibinfo{author}{\bibfnamefont{P.}~\bibnamefont{Cappellaro}},
  \bibinfo{author}{\bibfnamefont{L.}~\bibnamefont{Jiang}},
  \bibinfo{author}{\bibfnamefont{M.~V.} \bibnamefont{Dutt}},
  \bibinfo{author}{\bibfnamefont{E.}~\bibnamefont{Togan}},
  \bibinfo{author}{\bibfnamefont{A.~S.} \bibnamefont{Zibrov}},
  \bibnamefont{et~al.}, \bibinfo{journal}{Nature}
  \textbf{\bibinfo{volume}{455}}, \bibinfo{pages}{644} (\bibinfo{year}{2008}).

\bibitem[{\citenamefont{Balasubramanian
  et~al.}(2008)\citenamefont{Balasubramanian, Chan, Kolesov, Al-Hmoud, Tisler,
  Shin, Kim, Wojcik, Hemmer, Krueger et~al.}}]{Balasubramanian2008}
\bibinfo{author}{\bibfnamefont{G.}~\bibnamefont{Balasubramanian}},
  \bibinfo{author}{\bibfnamefont{I.~Y.} \bibnamefont{Chan}},
  \bibinfo{author}{\bibfnamefont{R.}~\bibnamefont{Kolesov}},
  \bibinfo{author}{\bibfnamefont{M.}~\bibnamefont{Al-Hmoud}},
  \bibinfo{author}{\bibfnamefont{J.}~\bibnamefont{Tisler}},
  \bibinfo{author}{\bibfnamefont{C.}~\bibnamefont{Shin}},
  \bibinfo{author}{\bibfnamefont{C.}~\bibnamefont{Kim}},
  \bibinfo{author}{\bibfnamefont{A.}~\bibnamefont{Wojcik}},
  \bibinfo{author}{\bibfnamefont{P.~R.} \bibnamefont{Hemmer}},
  \bibinfo{author}{\bibfnamefont{A.}~\bibnamefont{Krueger}},
  \bibnamefont{et~al.}, \bibinfo{journal}{Nature}
  \textbf{\bibinfo{volume}{455}}, \bibinfo{pages}{648} (\bibinfo{year}{2008}).

\bibitem[{\citenamefont{Dolde et~al.}(2011)\citenamefont{Dolde, Fedder,
  Doherty, Noebauer, Rempp, Balasubramanian, Wolf, Reinhard, Hollenberg,
  Jelezko et~al.}}]{Dolde2011}
\bibinfo{author}{\bibfnamefont{F.}~\bibnamefont{Dolde}},
  \bibinfo{author}{\bibfnamefont{H.}~\bibnamefont{Fedder}},
  \bibinfo{author}{\bibfnamefont{M.~W.} \bibnamefont{Doherty}},
  \bibinfo{author}{\bibfnamefont{T.}~\bibnamefont{Noebauer}},
  \bibinfo{author}{\bibfnamefont{F.}~\bibnamefont{Rempp}},
  \bibinfo{author}{\bibfnamefont{G.}~\bibnamefont{Balasubramanian}},
  \bibinfo{author}{\bibfnamefont{T.}~\bibnamefont{Wolf}},
  \bibinfo{author}{\bibfnamefont{F.}~\bibnamefont{Reinhard}},
  \bibinfo{author}{\bibfnamefont{L.~C.~L.} \bibnamefont{Hollenberg}},
  \bibinfo{author}{\bibfnamefont{F.}~\bibnamefont{Jelezko}},
  \bibnamefont{et~al.}, \bibinfo{journal}{Nature Physics}
  \textbf{\bibinfo{volume}{7}}, \bibinfo{pages}{459} (\bibinfo{year}{2011}),
  ISSN \bibinfo{issn}{1745-2473}, \eprint{1103.3432}.

\bibitem[{\citenamefont{Acosta et~al.}(2010)\citenamefont{Acosta, Bauch,
  Ledbetter, Waxman, Bouchard, and Budker}}]{Acosta2010}
\bibinfo{author}{\bibfnamefont{V.~M.} \bibnamefont{Acosta}},
  \bibinfo{author}{\bibfnamefont{E.}~\bibnamefont{Bauch}},
  \bibinfo{author}{\bibfnamefont{M.~P.} \bibnamefont{Ledbetter}},
  \bibinfo{author}{\bibfnamefont{A.}~\bibnamefont{Waxman}},
  \bibinfo{author}{\bibfnamefont{L.-S.} \bibnamefont{Bouchard}},
  \bibnamefont{and} \bibinfo{author}{\bibfnamefont{D.}~\bibnamefont{Budker}},
  \bibinfo{journal}{Physical Review Letters} \textbf{\bibinfo{volume}{104}},
  \bibinfo{pages}{070801} (\bibinfo{year}{2010}), ISSN
  \bibinfo{issn}{0031-9007}.

\bibitem[{\citenamefont{Cai et~al.}(2013)\citenamefont{Cai, Retzker, Jelezko,
  and Plenio}}]{Cai2013}
\bibinfo{author}{\bibfnamefont{J.}~\bibnamefont{Cai}},
  \bibinfo{author}{\bibfnamefont{a.}~\bibnamefont{Retzker}},
  \bibinfo{author}{\bibfnamefont{F.}~\bibnamefont{Jelezko}}, \bibnamefont{and}
  \bibinfo{author}{\bibfnamefont{M.~B.} \bibnamefont{Plenio}},
  \bibinfo{journal}{Nature Physics} \textbf{\bibinfo{volume}{9}},
  \bibinfo{pages}{1} (\bibinfo{year}{2013}).

\bibitem[{\citenamefont{Togan et~al.}(2010)\citenamefont{Togan, Chu, Trifonov,
  Jiang, Maze, Childress, Dutt, S{\o}rensen, Hemmer, Zibrov
  et~al.}}]{Togan2010}
\bibinfo{author}{\bibfnamefont{E.}~\bibnamefont{Togan}},
  \bibinfo{author}{\bibfnamefont{Y.}~\bibnamefont{Chu}},
  \bibinfo{author}{\bibfnamefont{A.~S.} \bibnamefont{Trifonov}},
  \bibinfo{author}{\bibfnamefont{L.}~\bibnamefont{Jiang}},
  \bibinfo{author}{\bibfnamefont{J.}~\bibnamefont{Maze}},
  \bibinfo{author}{\bibfnamefont{L.}~\bibnamefont{Childress}},
  \bibinfo{author}{\bibfnamefont{M.~V.~G.} \bibnamefont{Dutt}},
  \bibinfo{author}{\bibfnamefont{A.~S.} \bibnamefont{S{\o}rensen}},
  \bibinfo{author}{\bibfnamefont{P.~R.} \bibnamefont{Hemmer}},
  \bibinfo{author}{\bibfnamefont{A.~S.} \bibnamefont{Zibrov}},
  \bibnamefont{et~al.}, \bibinfo{journal}{Nature}
  \textbf{\bibinfo{volume}{466}}, \bibinfo{pages}{730} (\bibinfo{year}{2010}).

\bibitem[{\citenamefont{Bernien et~al.}(2012)\citenamefont{Bernien, Childress,
  Robledo, Markham, Twitchen, and Hanson}}]{Bernien2012}
\bibinfo{author}{\bibfnamefont{H.}~\bibnamefont{Bernien}},
  \bibinfo{author}{\bibfnamefont{L.}~\bibnamefont{Childress}},
  \bibinfo{author}{\bibfnamefont{L.}~\bibnamefont{Robledo}},
  \bibinfo{author}{\bibfnamefont{M.}~\bibnamefont{Markham}},
  \bibinfo{author}{\bibfnamefont{D.}~\bibnamefont{Twitchen}}, \bibnamefont{and}
  \bibinfo{author}{\bibfnamefont{R.}~\bibnamefont{Hanson}},
  \bibinfo{journal}{Phys. Rev. Lett.} \textbf{\bibinfo{volume}{108}},
  \bibinfo{pages}{043604} (\bibinfo{year}{2012}).

\bibitem[{\citenamefont{Sipahigil et~al.}(2012)\citenamefont{Sipahigil,
  Goldman, Togan, Chu, Markham, Twitchen, Zibrov, Kubanek, and
  Lukin}}]{Sipahigil2012}
\bibinfo{author}{\bibfnamefont{A.}~\bibnamefont{Sipahigil}},
  \bibinfo{author}{\bibfnamefont{M.~L.} \bibnamefont{Goldman}},
  \bibinfo{author}{\bibfnamefont{E.}~\bibnamefont{Togan}},
  \bibinfo{author}{\bibfnamefont{Y.}~\bibnamefont{Chu}},
  \bibinfo{author}{\bibfnamefont{M.}~\bibnamefont{Markham}},
  \bibinfo{author}{\bibfnamefont{D.~J.} \bibnamefont{Twitchen}},
  \bibinfo{author}{\bibfnamefont{A.~S.} \bibnamefont{Zibrov}},
  \bibinfo{author}{\bibfnamefont{A.}~\bibnamefont{Kubanek}}, \bibnamefont{and}
  \bibinfo{author}{\bibfnamefont{M.~D.} \bibnamefont{Lukin}},
  \bibinfo{journal}{Phys. Rev. Lett.} \textbf{\bibinfo{volume}{108}},
  \bibinfo{pages}{143601} (\bibinfo{year}{2012}).

\bibitem[{\citenamefont{Bernien et~al.}(2013)\citenamefont{Bernien, Hensen,
  Pfaff, Koolstra, Blok, Robledo, Taminiau, Markham, Twitchen, Childress
  et~al.}}]{Bernien2013}
\bibinfo{author}{\bibfnamefont{H.}~\bibnamefont{Bernien}},
  \bibinfo{author}{\bibfnamefont{B.}~\bibnamefont{Hensen}},
  \bibinfo{author}{\bibfnamefont{W.}~\bibnamefont{Pfaff}},
  \bibinfo{author}{\bibfnamefont{G.}~\bibnamefont{Koolstra}},
  \bibinfo{author}{\bibfnamefont{M.~S.} \bibnamefont{Blok}},
  \bibinfo{author}{\bibfnamefont{L.}~\bibnamefont{Robledo}},
  \bibinfo{author}{\bibfnamefont{T.~H.} \bibnamefont{Taminiau}},
  \bibinfo{author}{\bibfnamefont{M.}~\bibnamefont{Markham}},
  \bibinfo{author}{\bibfnamefont{D.~J.} \bibnamefont{Twitchen}},
  \bibinfo{author}{\bibfnamefont{L.}~\bibnamefont{Childress}},
  \bibnamefont{et~al.}, \bibinfo{journal}{Nature}
  \textbf{\bibinfo{volume}{497}}, \bibinfo{pages}{86} (\bibinfo{year}{2013}).

\bibitem[{\citenamefont{Hensen et~al.}(2015)\citenamefont{Hensen, Bernien,
  Dr{\'{e}}au, Reiserer, Kalb, Blok, Ruitenberg, Vermeulen, Schouten,
  Abell{\'{a}}n et~al.}}]{Hensen2015}
\bibinfo{author}{\bibfnamefont{B.}~\bibnamefont{Hensen}},
  \bibinfo{author}{\bibfnamefont{H.}~\bibnamefont{Bernien}},
  \bibinfo{author}{\bibfnamefont{A.~E.} \bibnamefont{Dr{\'{e}}au}},
  \bibinfo{author}{\bibfnamefont{A.}~\bibnamefont{Reiserer}},
  \bibinfo{author}{\bibfnamefont{N.}~\bibnamefont{Kalb}},
  \bibinfo{author}{\bibfnamefont{M.~S.} \bibnamefont{Blok}},
  \bibinfo{author}{\bibfnamefont{J.}~\bibnamefont{Ruitenberg}},
  \bibinfo{author}{\bibfnamefont{R.~F.~L.} \bibnamefont{Vermeulen}},
  \bibinfo{author}{\bibfnamefont{R.~N.} \bibnamefont{Schouten}},
  \bibinfo{author}{\bibfnamefont{C.}~\bibnamefont{Abell{\'{a}}n}},
  \bibnamefont{et~al.}, \bibinfo{journal}{Nature}
  \textbf{\bibinfo{volume}{526}}, \bibinfo{pages}{682} (\bibinfo{year}{2015}).

\bibitem[{\citenamefont{Huck et~al.}((2011))\citenamefont{Huck, Kumar, Shakoor,
  and Andersen}}]{Huck2011}
\bibinfo{author}{\bibfnamefont{A.}~\bibnamefont{Huck}},
  \bibinfo{author}{\bibfnamefont{S.}~\bibnamefont{Kumar}},
  \bibinfo{author}{\bibfnamefont{A.}~\bibnamefont{Shakoor}}, \bibnamefont{and}
  \bibinfo{author}{\bibfnamefont{U.~L.} \bibnamefont{Andersen}},
  \bibinfo{journal}{Phys. Rev. Lett.} \textbf{\bibinfo{volume}{106}},
  \bibinfo{pages}{096801} (\bibinfo{year}{(2011)}).

\bibitem[{\citenamefont{de~Leon et~al.}(2012)\citenamefont{de~Leon, Shields,
  Yu, Englund, Akimov, Lukin, and Park}}]{Leon2012}
\bibinfo{author}{\bibfnamefont{N.~P.} \bibnamefont{de~Leon}},
  \bibinfo{author}{\bibfnamefont{B.~J.} \bibnamefont{Shields}},
  \bibinfo{author}{\bibfnamefont{C.~L.} \bibnamefont{Yu}},
  \bibinfo{author}{\bibfnamefont{D.~E.} \bibnamefont{Englund}},
  \bibinfo{author}{\bibfnamefont{A.~V.} \bibnamefont{Akimov}},
  \bibinfo{author}{\bibfnamefont{M.~D.} \bibnamefont{Lukin}}, \bibnamefont{and}
  \bibinfo{author}{\bibfnamefont{H.}~\bibnamefont{Park}},
  \bibinfo{journal}{Phys. Rev. Lett.} \textbf{\bibinfo{volume}{108}},
  \bibinfo{pages}{226803} (\bibinfo{year}{2012}).

\bibitem[{\citenamefont{Riedel et~al.}(2017)\citenamefont{Riedel, S\"ollner,
  Shields, Starosielec, Appel, Neu, Maletinsky, and Warburton}}]{Riedel2017}
\bibinfo{author}{\bibfnamefont{D.}~\bibnamefont{Riedel}},
  \bibinfo{author}{\bibfnamefont{I.}~\bibnamefont{S\"ollner}},
  \bibinfo{author}{\bibfnamefont{B.~J.} \bibnamefont{Shields}},
  \bibinfo{author}{\bibfnamefont{S.}~\bibnamefont{Starosielec}},
  \bibinfo{author}{\bibfnamefont{P.}~\bibnamefont{Appel}},
  \bibinfo{author}{\bibfnamefont{E.}~\bibnamefont{Neu}},
  \bibinfo{author}{\bibfnamefont{P.}~\bibnamefont{Maletinsky}},
  \bibnamefont{and} \bibinfo{author}{\bibfnamefont{R.~J.}
  \bibnamefont{Warburton}}, \bibinfo{journal}{Phys. Rev. X}
  \textbf{\bibinfo{volume}{7}}, \bibinfo{pages}{031040} (\bibinfo{year}{2017}).

\bibitem[{\citenamefont{Hadden et~al.}(2010)\citenamefont{Hadden, Harrison,
  Stanley-Clarke, Marseglia, Ho, Patton, O'Brien, and Rarity}}]{Hadden2010}
\bibinfo{author}{\bibfnamefont{J.~P.} \bibnamefont{Hadden}},
  \bibinfo{author}{\bibfnamefont{J.~P.} \bibnamefont{Harrison}},
  \bibinfo{author}{\bibfnamefont{A.~C.} \bibnamefont{Stanley-Clarke}},
  \bibinfo{author}{\bibfnamefont{L.}~\bibnamefont{Marseglia}},
  \bibinfo{author}{\bibfnamefont{Y.~L.~D.} \bibnamefont{Ho}},
  \bibinfo{author}{\bibfnamefont{B.~R.} \bibnamefont{Patton}},
  \bibinfo{author}{\bibfnamefont{J.~L.} \bibnamefont{O'Brien}},
  \bibnamefont{and} \bibinfo{author}{\bibfnamefont{J.~G.}
  \bibnamefont{Rarity}}, \bibinfo{journal}{Applied Physics Letters}
  \textbf{\bibinfo{volume}{97}}, \bibinfo{pages}{0} (\bibinfo{year}{2010}).

\bibitem[{\citenamefont{Lukosz and Kunz}(1977)}]{Lukosz1977}
\bibinfo{author}{\bibfnamefont{W.}~\bibnamefont{Lukosz}} \bibnamefont{and}
  \bibinfo{author}{\bibfnamefont{R.~E.} \bibnamefont{Kunz}},
  \bibinfo{journal}{J. Opt. Soc. Am.} \textbf{\bibinfo{volume}{67}},
  \bibinfo{pages}{1607} (\bibinfo{year}{1977}).

\bibitem[{\citenamefont{Lukosz}(1979)}]{Lukosz1979}
\bibinfo{author}{\bibfnamefont{W.}~\bibnamefont{Lukosz}}, \bibinfo{journal}{J.
  Opt. Soc. Am.} \textbf{\bibinfo{volume}{69}}, \bibinfo{pages}{1495}
  (\bibinfo{year}{1979}).

\bibitem[{\citenamefont{Kaiser et~al.}(2009)\citenamefont{Kaiser, Jacques,
  Batalov, Siyushev, Jelezko, and Wrachtrup}}]{Kaiser2009}
\bibinfo{author}{\bibfnamefont{F.}~\bibnamefont{Kaiser}},
  \bibinfo{author}{\bibfnamefont{V.}~\bibnamefont{Jacques}},
  \bibinfo{author}{\bibfnamefont{A.}~\bibnamefont{Batalov}},
  \bibinfo{author}{\bibfnamefont{P.}~\bibnamefont{Siyushev}},
  \bibinfo{author}{\bibfnamefont{F.}~\bibnamefont{Jelezko}}, \bibnamefont{and}
  \bibinfo{author}{\bibfnamefont{J.}~\bibnamefont{Wrachtrup}}
  (\bibinfo{year}{2009}), \eprint{0906.3426},
  \urlprefix\url{http://arxiv.org/abs/0906.3426}.

\bibitem[{\citenamefont{Johnson and Christy}(1974)}]{Johnson1974}
\bibinfo{author}{\bibfnamefont{P.~B.} \bibnamefont{Johnson}} \bibnamefont{and}
  \bibinfo{author}{\bibfnamefont{R.~W.} \bibnamefont{Christy}},
  \bibinfo{journal}{Phys. Rev. B} \textbf{\bibinfo{volume}{9}},
  \bibinfo{pages}{5056} (\bibinfo{year}{1974}).

\end{thebibliography}

\end{document}